\documentclass[twocolumn]{aastex62} 
\usepackage{graphicx}
\usepackage{xcolor}
\usepackage{tabularx}
\usepackage{multirow}
\usepackage{amsmath}

\newcommand{\SectionSpace}{\vspace{10pt}}

\begin{document}

\title{
\textbf{ODIN: Comparison of Source Extractor and Tractor Source Detection for Lyman Alpha Emitting Galaxy Selection at $z$ = 3.1
}
}

\author[0009-0008-0197-8007]{Bailey Thompson}
\affiliation{Department of Physics and Astronomy, Rutgers, the State University of New Jersey, Piscataway, NJ 08854, USA}
\author[0000-0002-9811-2443]{Nicole Firestone}
\affiliation{Department of Physics and Astronomy, Rutgers, the State University of New Jersey, Piscataway, NJ 08854, USA}
\author[0000-0003-1530-8713]{Eric Gawiser}
\affiliation{Department of Physics and Astronomy, Rutgers, the State University of New Jersey, Piscataway, NJ 08854, USA}
\author[0000-0003-3004-9596]{Kyoung-Soo Lee}
\affiliation{Department of Physics and Astronomy, Purdue University, 525 Northwestern Ave., West Lafayette, IN 47906, USA}
\author[0000-0002-9176-7252]{Vandana Ramakrishnan}
\affiliation{Department of Physics and Astronomy, Purdue University, 525 Northwestern Ave., West Lafayette, IN 47906, USA}
\author[0000-0002-1172-0754]{Dustin Lang}
\affiliation{Perimeter Institute for Theoretical Physics, 31 Caroline Street North, Waterloo, ON N2L 2Y5, Canada}
\affiliation{Department of Physics and Astronomy, University of Waterloo, Waterloo, ON N2L 3G1, Canada}
\author[0000-0001-5567-1301]{Francisco Valdes}
\affiliation{National Optical Astronomy Observatory, 950 N. Cherry Avenue, Tucson, AZ 85719, USA}
\author[0000-0003-3428-7612]{Ho Seong Hwang}
\affiliation{Astronomy Program, Department of Physics and Astronomy,
Seoul National University, 1 Gwanak-ro, Gwanak-gu, Seoul 08826, Republic
of Korea}
\affiliation{SNU Astronomy Research Center, Seoul National University, 1
Gwanak-ro, Gwanak-gu, Seoul 08826, Republic of Korea}
\affiliation{Institute for Data Innovation in Science, Seoul National
University, Seoul 08826, Republic of Korea}
\author[0000-0002-4902-0075]{Lucia Guaita}
\affiliation{Universidad Andres Bello, Facultad de Ciencias Exactas, Departamento de Fisica y Astronomia, Instituto de Astrofisica, Fernandez Concha 700, Las Condes, Santiago RM, Chile}
\affiliation{Millennium Nucleus for Galaxies (MINGAL)}
\author[0000-0001-6270-3527]{Ankit Kumar}
\affiliation{Center for Cosmology and Computational Astrophysics, Institute for Advanced Study in Physics, Zhejiang University, Hangzhou 310027, China}
\affiliation{Institute of Astronomy, School of Physics, Zhejiang University, Hangzhou 310027, China}

\begin{abstract}
Lyman Alpha Emitting galaxies (LAEs) are young, actively star-forming galaxies identified by their strong Ly$\alpha$ emission at a rest-frame wavelength of 121.6\,nm\,$,$ making them valuable tracers of galaxy evolution and the large-scale structure of the Universe. The One-hundred-deg\textsuperscript{2} DECam Imaging in narrow-bands (ODIN) survey collects LAE samples during Cosmic Noon. Because LAE selection is based on photometric measurements derived from source-detection algorithms, the choice of detection method can directly affect the resulting LAE samples. We therefore compare two prominent source detection methods, Source Extractor (SE) and Tractor (Tr), to determine which is better suited for LAE selection in ODIN. We investigate differences between SE and Tr in the ODIN-COSMOS field, focusing on the $z=3.1$ LAE sample. We find that a significant fraction of Tr LAEs have low continuum flux measurements, increasing the inferred narrow-band excess and resulting in an LAE sample nearly twice as large as SE. However, LAEs detected by Tr but not SE exhibit equivalent width estimates indicative of potential contamination. We introduce an additional LAE selection criterion based on the expected Lyman-break behavior of LAEs at $z\sim3$ to reduce low-redshift contamination without requiring a significant detection of the break itself. Our results demonstrate that differences in source detection can significantly affect narrow-band LAE selection and motivate the use of SE photometry for ODIN LAE catalogs.

\vspace{25pt}
\end{abstract}

\section{Introduction}
\label{sec:Introduction}

Lyman Alpha Emitting Galaxies (LAEs) are star-forming galaxies as seen by their strong emission of Ly$\alpha$ at a rest-frame wavelength of 121.6\,nm \citep{Partridge1967}. Studies show that LAEs tend to be young and compact with low line-of-sight dust content \citep{Gawiser2007,Guaita2010,Pucha2022}, allowing Ly$\alpha$ photons to escape with less absorption and scattering before being observed. Therefore, LAEs can be efficiently identified and used to study the large-scale structure of the Universe and galaxy evolution over time \citep{Gawiser2007,Huang2022}.

Since the first discoveries of LAEs in the 1990s \citep{Hu1996,Pascarelle1996}, there have been a multitude of surveys focused on finding these high-$z$ galaxies \citep{Gawiser2006,DESI,Ouchi2018,Gebhardt2021,Lee2024}. These studies have revealed some differences in measured LAE properties; however, different definitions and LAE selection methods employed by these earlier studies may have contributed to these results, highlighting the need for a more systematic investigation.

Earlier studies such as the previously mentioned \citet{Hu1996} and \citet{Pascarelle1996}, as well as the Hawaii Survey \citep{Cowie1998}, used manual and visual inspection techniques to identify their LAEs because automated source detection had not yet been commonly utilized. Once the Large Area Lyman Alpha (LALA) survey \citep{Rhoads2000} employed Source Extractor (SE), automated source detection became standard in high-$z$ LAE surveys.

SE was released in 1996, the same year the first LAEs were discovered, and quickly gained traction due to its fast, flexible, and robust nature \citep{Bertin1996}. SE became one of the primary tools for source detection in high-$z$ galaxy surveys, including the Subaru Deep Field (SDF) survey \citep{Ouchi2003}, the Multiwavelength survey by Yale-Chile (MUSYC) \citep{Gawiser2007}, and the Lyman Alpha Galaxies in the Epoch of Reionization (LAGER) survey \citep{Zheng2017}. SE works by estimating the background level of an image, then uses pixel thresholding to detect and flag sources. After carefully deblending any merged objects, SE employs aperture-based photometry and returns a catalog of all sources and their resulting positions, shapes, fluxes, and quality flags.

The Tractor (Tr) is another prominent source detection method used in large multi-band imaging surveys \citep{Lang2016}. Released in 2016, Tr has gained wide-spread attention for its ability to handle complex large-area datasets and is now a core tool in surveys like the Dark Energy Spectroscopic Instrument (DESI) Legacy Survey \citep{DESI}. Tr measures photometry by forward modeling based on the local point spread function (PSF) for each source in an image. Model parameters are optimized to best match the image data, and forced photometry is applied to the other bands using positions and shapes derived from the detection image. This results in a catalog that contains PSF-aware flux measurements, as well as source positions, shapes, and associated uncertainties.

As SE and Tr become more widely used tools in the search for high-$z$ galaxies, the One-hundred-deg\textsuperscript{2} DECam Imaging in narrow-bands \citep[ODIN;][]{Lee2024} survey offers a valuable opportunity to directly compare these two source detection methods. ODIN is a NOIRLab survey which utilizes the Dark Energy Camera \citep[DECam;][]{Flaugher2015} on the 4m V\'ictor M. Blanco telescope at the Cerro Tololo Inter-American Observatory (CTIO) in Chile. With DECam, ODIN collects LAE samples at redshifts of 2.4, 3.1, and 4.5 using custom narrow-band filters that correspond to each redshift \citep[see][]{Firestone2024}. This range allows ODIN to look into the evolution of galaxies during what is known to be the peak epoch of star formation in the Universe, \textit{Cosmic Noon}. In addition to deep narrow-band images, ODIN uses archival \textit{grizy} broad-band images from the Hyper Supreme-Cam \citep[HSC;][]{Aihara2019} as well as \textit{u}-band images from the CFHT Large Area u-band Deep Survey \citep[CLAUDS;][]{Sawicki2019}. With this data, ODIN is expected to discover over 100,000 LAEs across 7 fields and 3 redshift bins, making it one of the largest LAE surveys to date \citep{Lee2024}.

Both SE and Tr are applicable to ODIN and future LAE surveys. In this paper, we analyze the difference between SE and Tr generated catalogs and their respective photometry for ODIN. Understanding the differences that arise from these selection methods will not only allow ODIN to create a more robust sample of LAEs, but it will also improve the accuracy of survey comparisons by clarifying methodological biases in source detection and measurement.

In Section \ref{sec:Data}, we summarize the ODIN imaging data, SE and Tr processing, and LAE selection methods. In Section \ref{sec:Results}, we present a comparison of the resulting catalogs, including photometric differences and evidence for potential contamination. In Section \ref{sec:Discussion}, we discuss the implications of these results for ODIN LAE selection and propose an additional LAE selection criterion, followed by conclusions and future work in Section \ref{sec:Conclusion}. In this work, we focus on selected LAEs for $z$ = 3.1 in the 9 deg\textsuperscript{2} ODIN-COSMOS field as this is ODIN's most well-studied redshift/field as of this publication. Throughout this work, we assume a $\Lambda$CDM cosmology with $h$ = 0.7, $\Omega_m$ = 0.27, and $\Omega_{\Lambda} = 0.73$. All magnitudes in this paper are in the AB magnitude system \citep{Oke1974}.

\SectionSpace


\section{Data}
\label{sec:Data}

\subsection{Images} 
\label{sec:Images}

ODIN combines $N501$ observations with broad-band images for LAE selection in the 9 deg\textsuperscript{2} COSMOS field. Both SE and Tr source catalogs utilize narrow-band data. ODIN's $N501$ filter was custom-made with a central wavelength of $501\pm3.8\,nm$ to detect LAEs at redshifts of 3.1, with a redshift interval of $\Delta z=0.059$ corresponding to the full-width-at-half-maximum (FWHM) of the filter transmission \citep{Lee2024}.

Archival \textit{grizy} broad-band images used in the SE and Tr source catalogs come from the second data release of the Hyper Supreme-Cam Subaru Strategic Program (HSC-SSP) on the 8.2\,m Subaru telescope located in Hawaii \citep{Aihara2019}. The HSC-SSP broad-band data include three layers in the COSMOS field: Wide, Deep, and Ultradeep. Both the SE and Tr catalogs have been created using a combination of the Deep and Ultradeep layers. In addition to archival \textit{grizy} broad-band images, both SE and Tr catalogs use \textit{u}-band images from The CFHT Large Area U-band Deep Survey (CLAUDS; \citeauthor{Sawicki2019}\,\citeyear{Sawicki2019}) using the MegaCam mosaic imager on the Canadian-France-Hawaii Telescope (CFHT). This \textit{u}-band data was created to match the depth of the Deep and Ultradeep layers in the COSMOS field with a median depth of 27.7\,mag \citep{Sawicki2019}.

\subsection{SE Source Catalog} 
\label{sec:SE Source Catalog}

For SE, the N501 detection image is a full-depth coadd of dithered DECam exposures processed with the NOIRLab DECam Community Pipeline \citep[CP;][]{Valdes2014,Valdes2021}. The COSMOS N501 coadd consists of 72 individual exposures and reaches a nearly uniform median depth of approximately 25.4 AB mag. At the individual-exposure level, the CP flags bad CCD pixels, saturated and bleed-trail pixels, cosmic rays, and other streaks, while satellite trails are additionally identified manually and masked. Flagged and masked pixels are excluded when constructing the final coadd. The broad-band measurement images consist of HSC-SSP coadded $grizy$ images and a CLAUDS $u$-band image. The HSC-SSP science images, data-quality maps, and RMS maps are reprojected using the DECam CP to match the tangent point and pixel scale of the ODIN data.

We run SE in dual-image mode, using the N501 coadd for detection and the $ugrizy$ broad-band images for measurement. All images are convolved with a 1 arcsecond FWHM Gaussian filter ({\tt FILTERING} = Y), approximately matched to the average narrow-band seeing. We adopt a detection threshold of $0.95\sigma$ (in units of the pre-convolution rms) and a minimum area of 1 pixel ({\tt DETECT\_THRESH} = 0.95; {\tt DETECT\_MINAREA} = 1), which maximize the detection of faint point sources while maintaining a low contamination rate \citep[$\lesssim 1$\%, see][for details]{Ramakrishnan2023}. Photometry is extracted in 12 apertures with diameters of 1, 1.5, 1.75, 2, 2.25, 2.75, 3, 3.5, 4, 4.5, 5, and 7 arcseconds.

We additionally impose a half-light radius cut ($R_{50}<1.38^{\prime\prime}$) and use two N501 ``half-stacks,'' each constructed from 36 exposures with the DECam CP, to identify remaining spurious sources. Sources are detected in the full-depth coadd and their fluxes are independently measured in each half-stack; strongly inconsistent measurements between the two are identified as likely spurious and removed (see \citet{Firestone2024} for further details).

\subsection{Tr Source Catalog} 
\label{sec:Tr Source Catalog}

Tr is run as part of the \texttt{Legacypipe} reduction pipeline \citep{legacypipe}.  Source detection, morphological type classification, and measurements of fluxes, positions, and galaxy shapes are performed on all overlapping images jointly.  The pipeline processes one ``brick'' of $0.25 \times 0.25$ deg$^2$ at a time. Source detection proceeds by merging detections in any of the three narrow-band filters, with detection performed as described in \citet{lang2026}.  Source fitting is carried out by nonlinear least-squares optimization: the code finds model parameters including source positions, fluxes, and galaxy shapes such that the PSF-convolved model predictions best match all the observed images that overlap.  Sources that have flux in only one of the narrow-band filters will end up with fit fluxes that are consistent with zero in the other bands.  After source positions and galaxy shapes have been measured in the narrow-band images, they are applied to the broad-band images via forced photometry: the source profiles are held fixed and only the broad-band fluxes are fit so that the PSF-convolved model predictions best match the observed images.

\subsection{Catalog Processing} 
\label{sec:Catalog Processing}

Our LAE samples are created by processing either the SE or Tr source catalog through a selection pipeline. For SE, we use the methodology described by \citet{Firestone2024}, while the Tr catalog is processed using the same pipeline adjusted to accommodate the differences in catalog structure. Both catalogs undergo multiple corrections and selection criteria, as described below, to guarantee robust LAE samples.

We first consider the scattering and absorption of light as it passes through dust in the Milky Way and makes its way to Earth. This effect causes sources to appear dimmer and redder than they truly are. Therefore, Galactic dust corrections are applied to sources found in both the SE and Tractor catalogs using the 8' resolution reddening map (SFD) from \citet{Schlegel1998} and Equation \ref{eq:Galactic dust correction};
\begin{equation}
\mathrm{flux}_{\mathrm{corr}} = \mathrm{flux}_{\mathrm{obs}} \times 10^{0.4kE(B-V)}.
\label{eq:Galactic dust correction}
\end{equation}
Here, $\mathrm{flux_{corr}}$ is the Galactic dust-corrected flux, $\mathrm{flux_{\mathrm{obs}}}$ is the observed flux, $k$ is the extinction coefficient for the corresponding filter, and $E(B-V)$ is the SFD reddening value for each source. The filter-specific extinction coefficients, $k$, are given in \citet{Firestone2024}. In addition, we assign a magnitude of 40 to sources with flux density values that have behavior indistinguishable from zero ($\leq 3.6 \times 10^{-7}~\mu\mathrm{Jy}$) to flag them as extremely faint objects. We choose to assign a magnitude of 40 to these objects because this magnitude does not overlap with the primary distribution of magnitudes, which peaks around 25\,mag.

When running the pipeline with our SE catalog, we also apply aperture corrections to improve the accuracy of our photometric measurements while limiting noise in our data. Our SE catalog includes flux density measurements for 12 different aperture diameters. As a compromise between noise and aperture loss, we use the 2.0 arcsecond aperture for each source's flux measurement. However, this approach underestimates the total flux density of our sources, since smaller apertures do not capture all of a source's light. To compensate, we compute the ratio of the median fractional flux in the chosen aperture (2.0 arcsecond) to the largest aperture (5.0 arcsecond) for bright, unsaturated point sources. We then use this fractional flux measurement as a correction factor to adjust the flux density upward to more accurately reflect a source's total brightness, while maintaining low levels of noise (for more details, see \citealt{Firestone2024}).

For both cases, we then apply star masking to exclude sources contaminated by bright stars. We use starmasks from the HSC-SSP \textit{g}-band images \citep{Coupon2018} to remove saturated pixels and regions affected by CCD blooming. This allows us to minimize contamination and maintain the reliability of our photometry. 

We then implement data quality cuts to ensure that our data are reliable and reduce contamination from low-quality detections that are not removed by star masking. We first require the narrow-band signal to noise ratio ($\mathrm{S}/\mathrm{N}_{\mathrm{NB}}$) to be above 5 in both SE and Tr. The remaining data quality cuts differ between SE and Tr catalogs due to their differences in diagnostic flags and metadata. SE quality cuts are primarily image-based and focus on the reliability of detections at a pixel level. We require \texttt{FLAGS}$<4$, allowing sources affected by neighboring objects or de-blending while excluding sources with more severe extraction issues, such as pixel saturation \citep{BA1996}. We additionally require \texttt{IMAFLAGS\_ISO}$=0$, indicating that none of the pixels associated with a source are flagged in the external flag map. We also ensure that flux measurements are non-zero to remove sources with unavailable or invalid photometry and include a half-light radius cut to exclude large, extended sources. In contrast, Tr quality cuts use both metadata from source model fitting and image-space quality flags. Tr uses the flag \texttt{BRICK\_PRIMARY} to ensure that objects are within the brick boundary, as well as \texttt{NOBS\_N501} to enforce a minimum number of observations for the narrow-band filter. Other quality criteria include the image-space flag \texttt{MASKBITS}, for which we require \texttt{MASKBITS} $=0$ to exclude sources associated with flagged regions, including bright and medium-bright Gaia stars, SGA galaxies, and saturated pixels. We additionally require \texttt{ALLMASK\_N501} $=0$ to specifically exclude sources affected by bad or saturated pixels in the $N501$ imaging. Similar image-quality filtering is applied in the SE catalog through the \texttt{IMAFLAGS\_ISO} and \texttt{FLAGS} cuts described above. Tr also produces a \texttt{FRACFLUX} metric for each catalog source, describing the fraction of flux contributed by neighboring sources within the fitting region and providing a measure of source blending. For the Tr catalog, we apply constraints on \texttt{FRACFLUX\_N501} to limit the contribution of flux from neighboring sources and help remove heavily blended objects with potentially unreliable photometry. We also require \texttt{FORCED\_MEAN\_FLUX\_IVAR\_BB}$>0$, where \texttt{FORCED\_MEAN\_FLUX\_IVAR\_BB} is the inverse variance associated with the forced mean broad-band flux measurement. This requirement ensures that sources have a valid broad-band flux measurement with an associated uncertainty, removing sources for which broad-band photometry is unavailable or invalid. A complete list of these cuts and thresholds is summarized in Table \ref{table data quality cuts}.

We note that the SE and Tr catalogs cover the same area within the ODIN-COSMOS field, allowing their source populations to be directly compared without normalization for differences in survey footprint.


\begin{table}[ht]
\centering
\small
\begin{tabular}{ll}
\hline
SE & Tr \\
\hline
$\mathrm{S}/\mathrm{N}_{\mathrm{NB}} \geq 5$ &
$\mathrm{S}/\mathrm{N}_{\mathrm{NB}} \geq 5$ \\
$\mathtt{IMAFLAGS\_ISO} = 0$ &
$\mathtt{BRICK\_PRIMARY} = \mathrm{True}$ \\
$\mathtt{FLAGS} < 4$ &
$\mathtt{MASKBITS} = 0$ \\
$f_\nu \ne 0$ &
$\mathtt{ALLMASK\_N501} = 0$ \\
$R_{50} < 1.38^{\prime\prime}$ &
$\mathtt{NOBS\_N501} > 10$ \\
&
$0 \leq \mathtt{FRACFLUX\_N501} \leq 0.5$ \\
&
$\mathtt{FORCED\_MEAN\_FLUX\_IVAR\_BB} > 0$ \\
\hline
\end{tabular}
\caption{Data quality cuts applied to the SE and Tractor catalogs. See Section \ref{sec:Catalog Processing} for a detailed description of each cut.}
\label{table data quality cuts}
\end{table}

\subsection{LAE Selection} 
\label{sec:LAE Selection}

The final step in our pipeline is to compile our LAE sample based on specific selection criteria. The criteria for LAE classification primarily rely on the \textit{hybrid-weighted double-broad-band continuum estimation} technique introduced in \citet{Firestone2024}. This technique estimates the continuum emission at the effective wavelength of the narrow-band filter by combining measurements from the two broad-band filters that bracket it. For $N501$, the $g$ and $r$-band magnitudes are linearly weighted to construct a synthetic continuum magnitude, denoted $gr$, following
\begin{equation}
gr = 0.856g + 0.144r.
\label{eq:gr_continuum}
\end{equation}
This continuum estimate is then compared to the narrow-band magnitude to compute the narrow-band excess ($gr-N501$), which quantifies the strength of Ly$\alpha$ emission above the underlying continuum. If the narrow-band excess of a source exceeds the minimum narrow-band excess cut,
\begin{equation}
gr-N501 > 0.83\,\mathrm{mag}
\label{eq:nbex_cut}
\end{equation}
(corresponding to a rest-frame Ly$\alpha$ equivalent width of 20\,\AA), and has a significance greater than $3\sigma$,
the source may be classified as an LAE candidate. We also require sources to have a broadband flux consistent with the limit expected for a pure emission-line (infinite-EW) object, thereby reducing contamination from spurious narrow-band detections. In addition to these criteria, we require our LAEs to have a narrow-band magnitude $\geq 20$\,mag and be brighter than the median $5\sigma$ depth of the narrow-band image to remove likely interlopers.

Once we apply Galactic dust extinction corrections, aperture corrections, star masking, data quality cuts, and our LAE selection criteria, we are left with two clean samples of LAEs for study; the SE LAE catalog, containing 5,354 sources, and the Tr LAE catalog, containing 12,521 sources.

\SectionSpace


\section{Results}
\label{sec:Results}

\subsection{Photometric Differences} 
\label{sec:Photometric Differences}

To understand the origin of the differences in LAE selection between SE and Tr, we first examine how the two methods measure photometry for the source catalogs created in Section \ref{sec:Catalog Processing} i.e., before LAE selection. By comparing flux measurements across all detected objects in the narrow-band and broad-band filters, we isolate systematic differences in flux estimation and uncertainty modeling that are present at the source catalog level. Examining these differences provides essential context for interpreting LAE selection behavior observed between the two detection methods.

We separate the \textit{SE Source Catalog} and \textit{Tr Source Catalog} into three sub-samples: sources detected by both SE and Tr (70\% of the \textit{SE Source Catalog}, 43\% of the \textit{Tr Source Catalog}), sources detected only by SE (30\% of the \textit{SE Source Catalog}), and sources detected only by Tr (57\% of the \textit{Tr Source Catalog}). Direct photometric comparisons are possible only for sources detected by both SE and Tr, while sources detected only by either SE or Tr are discussed separately. For median flux values of each of these categories, see Table \ref{tab:full photo}.

\begin{figure*}[t!]
    \centering
    \includegraphics[width=\linewidth]{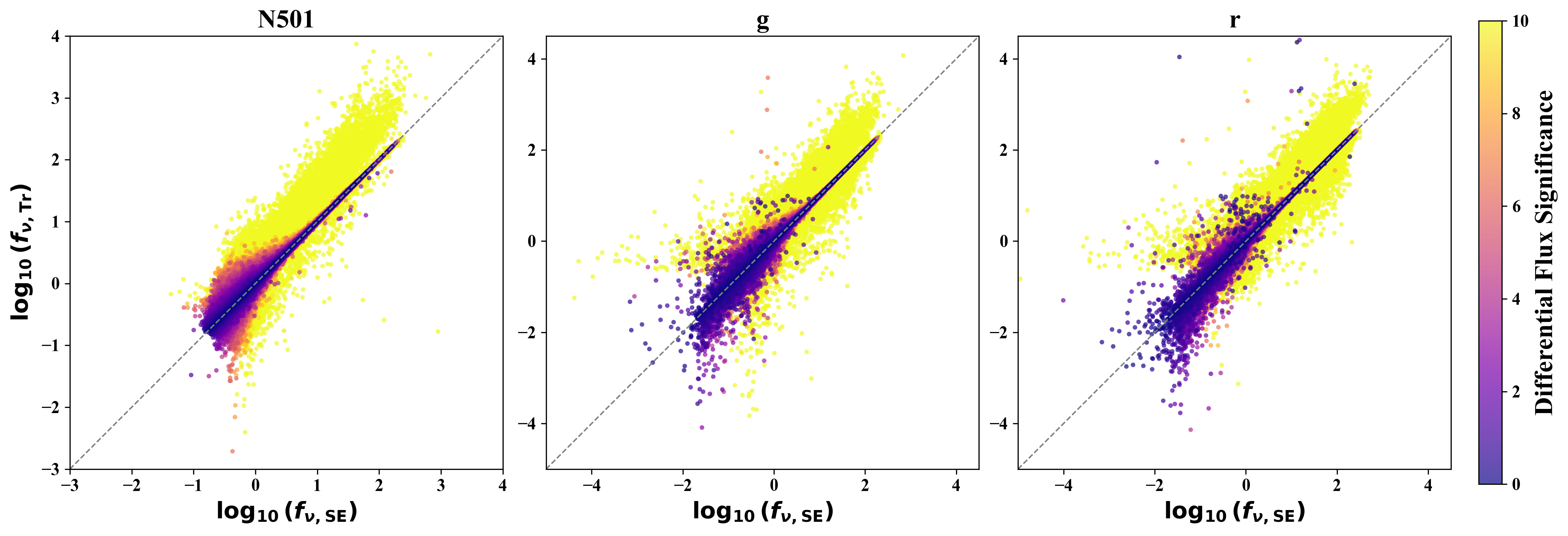}
    \caption{Comparison of SE and Tr flux density measurements in the $N501$, $g$, and $r$ filters for sources detected by both SE and Tr. Each panel shows SE versus Tr measured flux densities in $\mu$Jy, with the gray dashed line indicating the 1:1 relation. Points are colored by the differential flux significance, i.e. the difference divided by its uncertainty, highlighting statistically significant discrepancies between the two methods.}
    \label{fig:Full Flux}
\end{figure*}

For sources detected by both SE and Tr, we compare their flux density measurements in the narrow-band and broad-band filters to assess the level of agreement between the two methods. In particular, we focus on the $N501$ narrow-band and the $g$ and $r$ bands, as these filters are used to measure narrow-band excess, which is critical in our selection of LAEs. Figure \ref{fig:Full Flux} shows SE versus Tr flux density measurements for sources detected by both, with points colored by differential flux significance,

\begin{equation}
S_{\Delta f} =
\frac{\left|f_{\nu,\mathrm{SE}} - f_{\nu,\mathrm{Tr}}\right|}
{\sqrt{\sigma_{f_{\nu,\mathrm{SE}}}^{2} +
\sigma_{f_{\nu,\mathrm{Tr}}}^{2}}}.
\label{eq:diff flux significance}
\end{equation}

Because SE and Tr measure fluxes from the same underlying imaging, their measurements are not statistically independent. We therefore use $S_{\Delta f}$ as a metric for comparing flux differences relative to the reported uncertainties rather than as a formal statistical significance. This representation highlights both the overall agreement between the pipelines and the subset of sources for which the measured fluxes differ substantially relative to their uncertainties.

We find good agreement between SE and Tr flux density measurements across all filters, with the majority of sources lying close to the 1:1 relation, particularly at low differential flux significance and high flux. In the $N501$ narrow-band, most sources cluster tightly around the 1:1 line, though a population of objects show significant deviations at higher fluxes. In this regime, Tr tends to measure higher flux densities than SE, indicating a systematic, method-dependent offset at high measured flux densities. We observe similar trends in the $g$ and $r$ broad-bands, along with more scatter and pronounced asymmetric departures from the 1:1 relation, consistent with stronger method-dependent differences in broad-band flux estimation.

To further investigate these photometric differences between SE and Tr, we now compare sources detected only by one of the two methods. Similar to SE and Tr sources, we focus on the $N501$, $g$, and $r$ filters and present the median flux values of each filter and category in Table \ref{tab:full photo}.

For sources detected only by Tr, we have found that the median flux density values across all filters are lower than those of the sources detected by both SE and Tr. For sources detected only by SE, the median $N501$ flux density is higher than that of sources detected by both SE and Tr, while the median flux densities in all broad-band filters are consistent with zero. Only 9\% of SE-only sources exhibit positive $g$-band flux, indicating extremely weak or undetected continuum emission. This behavior suggests that SE preferentially recovers narrow-band-dominated sources with very weak continuum, likely reflecting a higher vulnerability to narrow-band-only spurious detections. The majority of these objects are removed by our current LAE selection. However, a small fraction remains in the final LAE samples. We address these remaining contaminants by introducing an additional LAE selection criterion in Section \ref{sec:New LAE Selection Criterion}.

\begin{table}[ht]
\centering
\begin{tabular}{l|cc|cc}
\hline
& \multicolumn{2}{c|}{SE Photometry} & \multicolumn{2}{c}{Tr Photometry} \\
\cline{2-5}
& SE Only & SE and Tr & Tr Only & SE and Tr \\
\hline\hline
$N501$ & 0.986 & 0.619 & 0.179 & 0.715 \\
$g$-band & 0 & 0.511 & 0.140 & 0.573 \\
$r$-band & 0 & 0.762 & 0.216 & 0.849 \\
\hline
\end{tabular}
\caption{Median flux densities measured by SE and Tr in the $N501$, $g$, and $r$ filters for each source category, with all values reported in units of $\mu$Jy.}
\label{tab:full photo}
\end{table}

\subsection{Categories of LAEs} 
\label{sec:Categories of LAEs}

\begin{figure*}[t!]
    \centering
    \includegraphics[width=\linewidth]{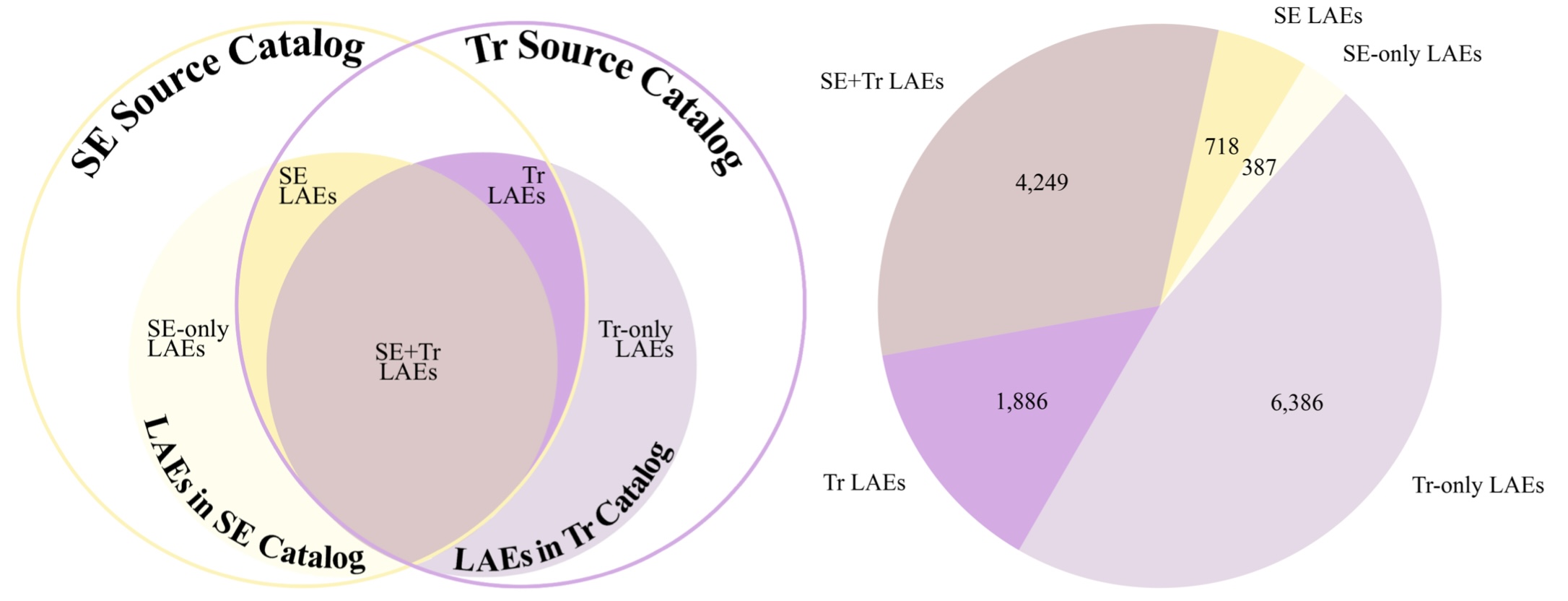}
    \caption{The left panel is a Venn diagram describing the five categories used in this analysis. The smaller filled circles represent the \textit{LAEs in SE Source Catalog} (yellow) and the \textit{LAEs in Tr Source Catalog} (purple), while the larger outlined circles represent the full \textit{SE Source Catalog} and \textit{Tr Source Catalog}. The overlapping regions describe the five categories: \textit{SE-only LAEs}, \textit{Tr-only LAEs}, \textit{SE LAEs}, \textit{Tr LAEs}, and \textit{SE+Tr LAEs}. The right panel is a pie chart that shows how ODIN-COSMOS LAEs are distributed across those five categories.}
    \label{fig:VDPC Plots}
\end{figure*}

 We then separate our SE and Tr LAE samples into five different categories. Comparing the samples in these categories allows us to understand the differences between SE and Tr source detection methods.

We separate our SE and Tr LAE catalogs according to how they were identified in SE and Tr to investigate the systematic differences between the two methods. The five categories include:
\begin{itemize}
    \item[$\circ$] \textit{SE-only LAEs}: Objects missing from the Tr catalog but selected as LAEs in the SE catalog.
    \item[$\circ$] \textit{Tr-only LAEs}: Objects missing from the SE catalog but selected as LAEs in the Tr catalog.
    \item[$\circ$] \textit{SE LAEs}: In both the SE and Tr source catalogs but selected as LAEs only in SE.
    \item[$\circ$] \textit{Tr LAEs}: In both the SE and Tr source catalogs but selected as LAEs only in Tr.
    \item[$\circ$] \textit{SE+Tr LAEs}: In both the SE and Tr source catalogs and selected as LAEs in both catalogs.
\end{itemize}
The LAEs in these categories come from the \textit{LAEs in SE Source Catalog} or the \textit{LAEs in Tr Source Catalog}, and are assigned to a category depending on if they are also found in the broader \textit{SE Source Catalog}, \textit{Tr Source Catalog}, or both. To illustrate how each category is derived, we present a Venn diagram as well as a pie chart with relative population sizes for each category in Figure \ref{fig:VDPC Plots}.

The \textit{SE-only LAEs} category contains the smallest population of objects, comprising only 3\% of the total LAEs included in this analysis. \textit{SE-only LAEs} are those found exclusively in the \textit{SE Source Catalog} with photometric measurements available only from SE. The \textit{Tr-only LAEs} represent a much greater portion, which constitutes about 47\% of the LAEs analyzed. These objects appear only in the \textit{Tr Source Catalog}, and therefore only have data derived from Tr. 

The \textit{SE LAEs} and \textit{Tr LAEs} categories yield 5\% and 14\%, respectively, of the total LAEs. Finally, the \textit{SE+Tr LAEs} category makes up a significant portion of the LAEs used in this analysis, accounting for 31\% of the sample. Objects in these last three categories have photometric data from both source detection methods.

By examining these five categories, we perform a detailed analysis of the differences between LAEs selected by SE and Tr in the COSMOS field at $z$ = 3.1. Each category offers a distinct perspective on the systematic discrepancies and photometric uncertainties introduced by the differing source detection methods.

\subsection{LAE Selection Cuts} 
\label{sec:Omitted LAEs}

For our LAE categories, we first examine why \textit{SE LAEs} and \textit{Tr LAEs} were not selected as LAEs by the other detection method. These two categories are unique in that they contain objects with photometry measured by both SE and Tr while being classified as an LAE by only one of the detection methods, making them the only samples for which we can directly evaluate the reason they do not pass our selection criteria in one method.

For each object in these categories, we test whether it satisfies each individual selection criterion described in Section \ref{sec:LAE Selection} to better understand why they were not classified as LAEs by both SE and Tr. Because objects may fail multiple criteria simultaneously, Table \ref{LAE Cuts Table} reports the fraction of objects in each category that fail each criterion when considered individually, regardless of whether they also fail other cuts. We find that the narrow-band excess (NB Excess) and narrow-band excess significance (NB Excess 3$\sigma$) cuts dominate the differences in LAE classification between SE and Tr, while all other criteria affect only a small fraction of sources.


\begin{table}[ht]
\centering
\small
\begin{tabular}{@{}lr@{\hspace{1.5em}}lr@{}}
\hline
\multicolumn{2}{c}{SE LAEs} &
\multicolumn{2}{c}{Tr LAEs} \\
\hline\hline
Primary: & 0.0\% & zero flux: & 0.0\% \\
Maskbits: & 3.5\% & flags: & 0.5\% \\
Allmask: & 0.0\% & imaflags: & 1.5\% \\
NOBS: & 20.5\% & S/N: & 1.0\% \\
Fractional Flux: & 8.1\% & & \\
S/N: & 3.2\% & & \\
Ivar: & 4.7\% & & \\
\hline
NB Excess: & 50.8\% & NB Excess: & 79.2\% \\
NB Excess 3$\sigma$: & 32.9\% &
NB Excess 3$\sigma$: & 48.6\% \\
Inf EW: & 0.8\% & Inf EW: & 0.3\% \\
NB Mag: & 0.0\% & NB Mag: & 0.0\% \\
NB Depth: & 1.5\% & Flux Radius: & 3.9\% \\
& & NB Depth: & 0.5\% \\
& & Halfstacks: & 4.8\% \\
\hline
\end{tabular}
\caption{Percentages of objects in the \textit{SE LAEs} and \textit{Tr LAEs} categories that fail each selection criterion when evaluated individually. \textit{SE LAEs} are evaluated using Tr photometry and \textit{Tr LAEs} using SE photometry. The horizontal line separates the data quality criteria and LAE criteria as discussed in Section \ref{sec:LAE Selection}.}
\label{LAE Cuts Table}
\end{table}

We also examine the overlap between objects rejected by the NB Excess and NB Excess 3$\sigma$ criteria. Among \textit{SE LAEs}, 11\% of sources fail both cuts, while 40\% fail only the NB Excess cut and 22\% fail only the NB Excess 3$\sigma$ cut. In contrast, for \textit{Tr LAEs}, 43\% of objects fail both criteria, with 36\% failing only the NB Excess cut and just 6\% failing only the NB Excess 3$\sigma$ cut. This asymmetric failure pattern indicates that the two pipelines reject each other's LAE candidates for somewhat different reasons. Tr's rejection of \textit{SE LAEs} is split between objects that fail due to narrow-band excess and those that fail due to narrow-band excess significance, perhaps reflecting how Tr’s forward-modeling affects fluxes and uncertainties in different ways for different sources. In contrast, SE's rejection of \textit{Tr LAEs} is primarily the result of reduced narrow-band excess measurements, with narrow-band excess significance failures largely accompanying the weaker excess rather than acting as an independent source of rejection.

\subsection{Narrow-band Excess} 
\label{sec:narrow-band Excess}

\begin{figure*}[t!]
    \centering
    \includegraphics[width=\linewidth]{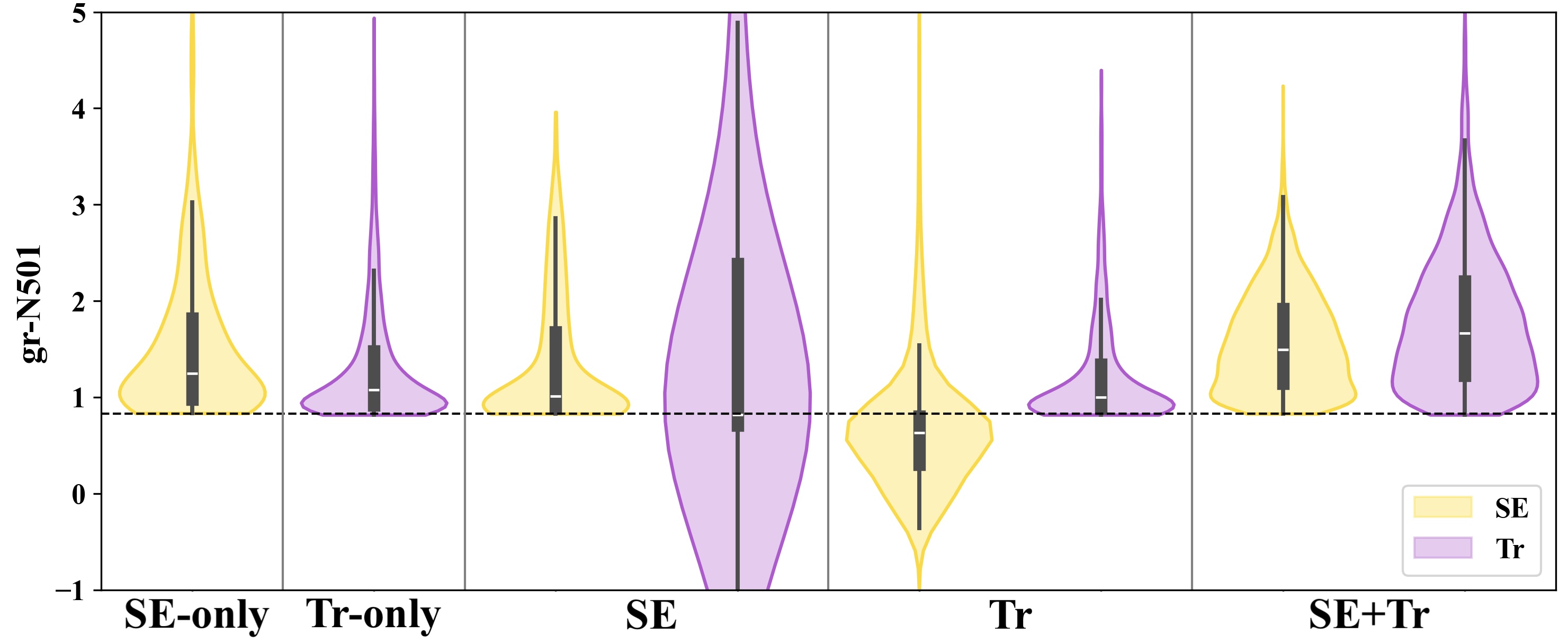}
    \caption{Distributions of narrow-band excess $-$ the difference between the underlying continuum and Ly$\alpha$ emission of an object ($gr-N501$) $-$ for each LAE category. Categories are separated by the gray vertical lines, with SE (yellow, left) and Tr (purple, right) measurements shown for each category. The boxplots within each distribution indicate the median and interquartile range (25th-75th percentiles). The dashed horizontal line marks the narrow-band excess selection threshold of 0.83 mags.}
    \label{fig:NB Ex}
\end{figure*}

Narrow-band excess, defined as the magnitude difference between the continuum and Ly$\alpha$ emission ($gr-N501$), is one of the most important criteria for identifying an LAE. A narrow-band excess greater than 0.83 mags indicates that the Ly$\alpha$ emission is sufficiently strong to be distinguished from the continuum, making the source a potential LAE candidate.
We have also found in Section \ref{sec:Omitted LAEs} that our narrow-band excess criterion is the most prominent cut in our selection. Therefore, we analyze the different narrow-band excess distributions in each LAE category for both SE and Tr measurements as well as the potential reasons behind the differences.

Figure \ref{fig:NB Ex} shows the distributions of narrow-band excess measurements for each LAE category, with the \textit{SE LAEs}, \textit{Tr LAEs}, and \textit{SE+Tr LAEs} illustrating the measurement differences between SE and Tr for the same set of objects. For the \textit{SE+Tr LAEs}, we find that both the median narrow-band excess and the overall distribution are higher in Tr, and an object-by-object comparison shows that Tr generally measures a higher narrow-band excess than SE. This trend can also be seen in the \textit{Tr LAEs} category, with many Tr-measured objects just above our narrow-band excess cut while their SE measurement counterparts primarily sit below, aligning with our results from Section \ref{sec:Omitted LAEs}. The \textit{SE LAEs} category exhibit a more skewed distribution in their Tr measurements, with roughly half of the objects lying below the selection cut, while the minimum and maximum measurements extend to substantially more extreme narrow-band excesses than seen in the other LAE categories. Finally, although the \textit{SE-only LAEs} and \textit{Tr-only LAEs} do not allow for direct cross-method comparison, their median narrow-band excess values fall between the corresponding detection method medians of LAEs they identify in other categories: \textit{SE-only LAEs} lie between the SE measurements of the \textit{SE LAEs} and \textit{SE+Tr LAEs}, and \textit{Tr-only LAEs} lie between the Tractor measurements of the \textit{Tr LAEs} and \textit{SE+Tr LAEs}.

These results indicate that differences between SE and Tr LAE selections are strongly influenced by systematic offsets in measured narrow-band excess. Tr measurements are generally higher, leading many \textit{Tr LAEs} to lie just above the narrow-band excess cut while their SE counterparts fall below. At the same time, the presence of highly skewed and extreme Tr measurements in the \textit{SE LAEs} demonstrates that method-dependent outliers also contribute to selection differences. Given that the narrow-band excess is defined by the relative difference between continuum and narrow-band measurements, we next investigate which component of the photometry drives this systematic difference.

\begin{figure}[t!]
    \centering
    \includegraphics[width=\linewidth]{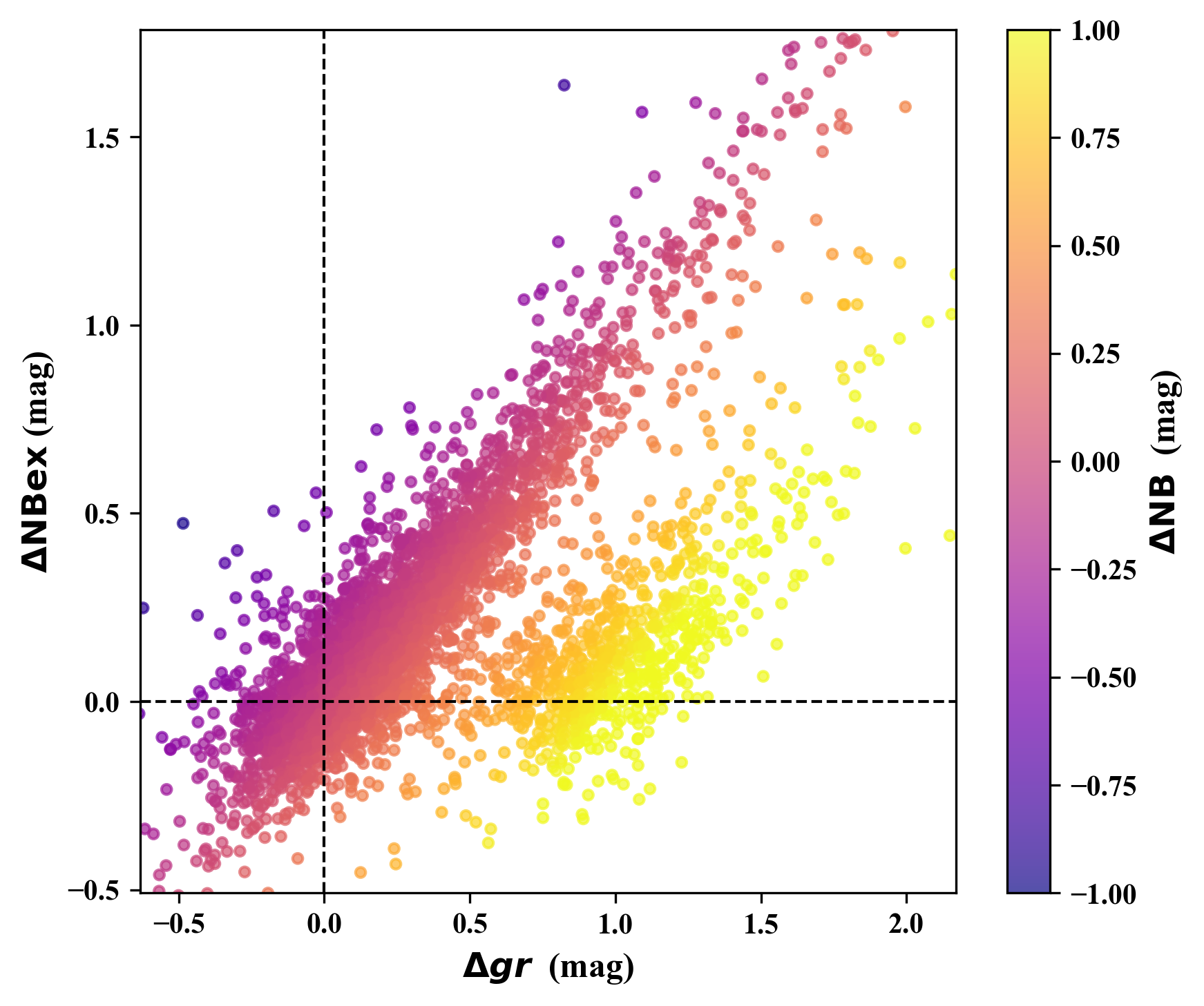}
    \caption{Difference in narrow-band excess between Tr and SE ($\Delta$NBex) versus the difference in estimated continuum magnitude ($\Delta$$gr$) for \textit{SE+Tr LAEs}. Points are colored by the corresponding difference in narrow-band magnitude ($\Delta$NB). The dashed lines indicate the zero-difference crossover, with positive values corresponding to fainter fluxes in Tr, while negative values correspond to fainter fluxes in SE.}
    \label{fig:NBex AB}
\end{figure}

In Figure \ref{fig:NBex AB} we present the difference in narrow-band excess between SE and Tr ($\Delta$NBex) as a function of the difference in estimated continuum magnitude ($\Delta gr$) for \textit{SE+Tr LAEs}. Both differences are calculated as the Tr measurement minus the corresponding SE measurement. Points are colored by the difference in narrow-band magnitude ($\Delta$NB), calculated in the same manner.

A strong, approximately linear correlation is observed between $\Delta$NBex and $\Delta$$gr$, indicating that systematic differences in continuum estimation dominate the observed narrow-band excess offsets between SE and Tr. The color gradient reveals that while narrow-band magnitudes generally track continuum differences between the two methods, they do not exhibit a strong correlation with the narrow-band excess itself. These trends can also be seen for \textit{SE LAEs} and \textit{Tr LAEs}, and overall, these results demonstrate that differences in narrow-band excess measurements are primarily driven by method-dependent differences in the broad-band photometry used to estimate the continuum.

\subsection{Equivalent Width} 
\label{sec:Equivalent Width}

\begin{figure*}[t!]
    \centering
    \includegraphics[width=\linewidth]{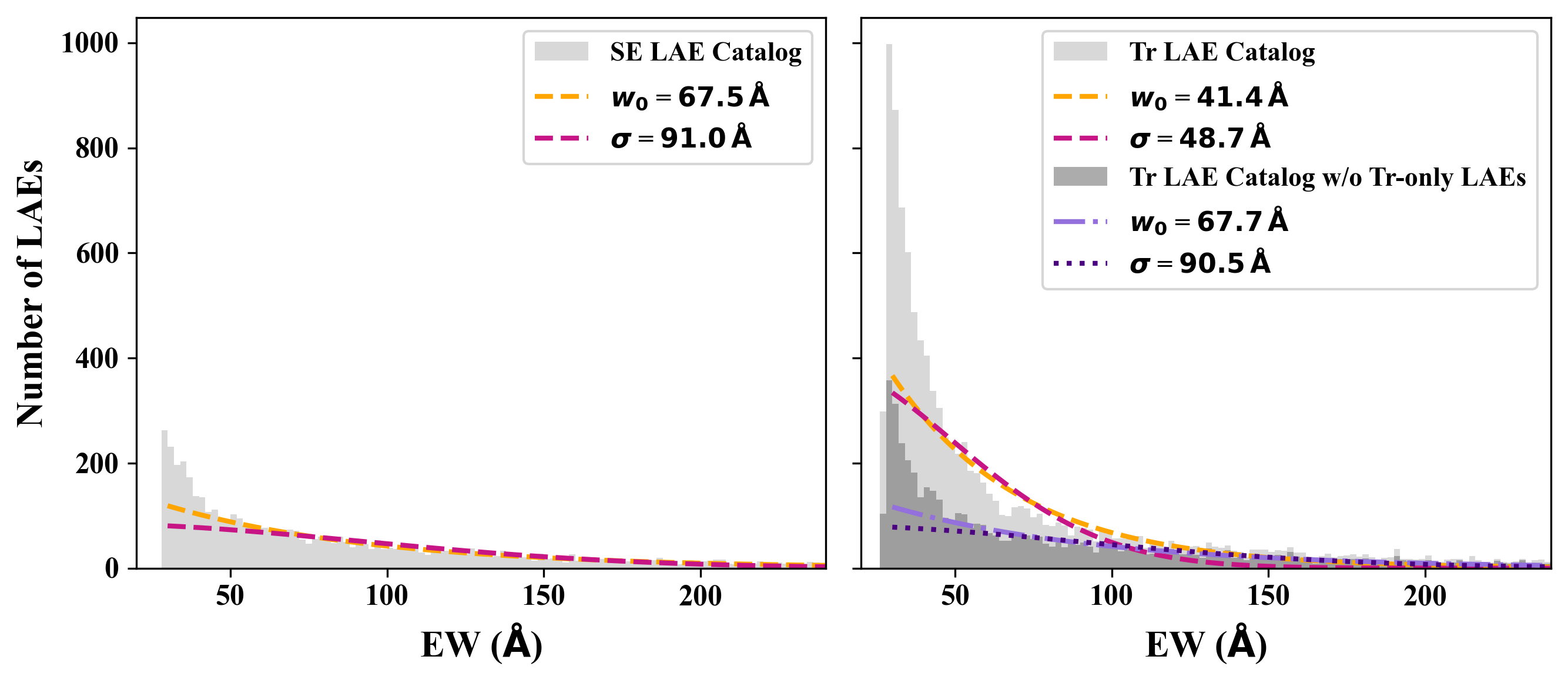}
    \caption{Rest-frame equivalent width (EW) distributions and fits for the \textit{SE LAE Catalog} (left) and \textit{Tr LAE Catalog} (right). The x-axis represents EW measurements in Angstroms. The y-axis represents the number of LAEs in each EW bin.  The dashed orange and magenta curves show the best-fitting exponential ($w_0$) and Gaussian ($\sigma_{\textrm{gauss}}$) models, respectively. The right panel additionally shows the EW distribution of the Tr LAE catalog after removing the \textit{Tr-only LAEs} (dark gray), together with the corresponding best-fitting exponential (purple dash-dotted) and Gaussian (purple dotted) models. The sharp cut off in EW at the low-EW end corresponds to the minimum narrow-band excess selection threshold.}
    \label{fig:EW Dist}
\end{figure*}

Differences in the photometric measurements produced by SE and Tr can affect not only which sources satisfy our LAE selection criteria, but also the inferred properties of the resulting LAE samples. Comparing the equivalent-width distributions of the SE and Tr LAE samples therefore provides an additional way to assess how differences between the two photometric methods influence LAE selection and the properties of the resulting catalogs.

The equivalent width (EW) of a spectral line quantifies its strength relative to the underlying continuum. It is defined as the width of a rectangle, with height equal to the continuum level, that has the same integrated area as the spectral feature. Therefore, we use EW to characterize our LAE samples, as it provides a measure of Ly$\alpha$ emission strength relative to the continuum and offers insight into the underlying galaxy properties.
For our EW distributions, we follow the methodologies of \citet{Venemans2005} and \citet{Guaita2010}. We define the observed rest-frame Ly$\alpha$ equivalent width as 
\begin{equation}
EW_{\textrm{obs}} = {A}/{B},
\label{eq:ew obs}
\end{equation}
where
\begin{equation}
A = Q_{\mathrm{NB}} - Q_{ab} \, 10^{(ab - NB)/2.5}
\label{eq:A}
\end{equation}
and
\begin{equation}
\begin{aligned}
B =
\frac{
w_{\textrm{BB}} \, T_{\textrm{EL,BB}}\!\left({c}/{\lambda_{\textrm{EL}}^2}\right) 10^{(ab - NB)/2.5}
}{
\int \left({c}/{\lambda^2}\right) T_{\textrm{BB}}(\lambda)\, d\lambda
}
\\
-
\frac{
T_{\textrm{EL,NB}}\!\left({c}/{\lambda_{\textrm{EL}}^2}\right)
}{
\int \left({c}/{\lambda^2}\right) T_{\textrm{NB}}(\lambda)\, d\lambda
}.
\end{aligned}
\label{eq:B}
\end{equation}
In Equation~\ref{eq:A}, $Q_{\mathrm{NB}}$ and $Q_{ab}$ represent the fraction of continuum flux transmitted after absorption by the Ly$\alpha$ forest in the narrow-band ($N501$) and the double-broad-band continuum estimation ($gr$), respectively. We compute $Q$ for a given filter as
\begin{equation}
Q_{\mathrm{filter}} =
\frac{
\int e^{-\tau_{\textrm{eff}}(\lambda)} \left({c}/{\lambda^2}\right) T(\lambda)\, d\lambda
}{
\int \left({c}/{\lambda^2}\right) T(\lambda)\, d\lambda
},
\label{eq:Q}
\end{equation}
where $c$ is the speed of light and $T(\lambda)$ is the filter transmission as a function of wavelength. $\tau_{eff}(\lambda)$ is the effective optical depth resulting from the attenuation of continuum emission blueward of Ly$\alpha$ \citep[ie. the Ly$\alpha$ forest][]{Press1993, Madau1995, Venemans2005},
as defined below.
\begin{equation}
\tau_{\textrm{eff}}(\lambda) = 0.0036 \left( \frac{\lambda}{1216\,\text{\AA}} \right)^{3.46}
\label{eq:tau}
\end{equation}

In Equation \ref{eq:B}, $w_{\textrm{BB}}$ is the weight assigned to the broad-band, BB, in which the Ly$\alpha$ emission line resides. For $z$ = 3.1 LAEs, this broad-band corresponds to the $g$-band. Because the exact Ly$\alpha$ wavelength is not known for individual LAEs, we use the narrow-band transmission curve as a proxy for the expected redshift distribution. DESI spectroscopic confirmations of a subset of ODIN LAEs have revealed that this is a realistic assumption (see \citet{white2024clustering}). We then compute $T_{\textrm{EL,filt}}$ as a weighted average of the filter transmission over this distribution. $\lambda_{\textrm{EL}}$ is the wavelength of the emission line, which is set to 501 nm, the effective wavelength of the $N501$ narrow-band filter \citep{Lee2024}.

We then fit the EW distributions to characterize the overall distribution of equivalent widths and to determine whether the LAE samples selected by SE and Tractor exhibit systematic differences. Previous studies have found that LAE EW distributions are reasonably described by exponential or Gaussian functional forms \citep[e.g.][]{Gronwall2007, Ouchi2008, Nilsson2009, Guaita2010, Zheng2012, Kerutt2022}. We therefore use these distributions as empirical models for the expected shape of the LAE EW distribution. We fit both an exponential distribution (Equation \ref{eq:exp}) and a Gaussian distribution (Equation \ref{eq:gauss}), as shown below.

\begin{equation}
N = C \exp\left(-\frac{EW}{w_0}\right)
\label{eq:exp}
\end{equation}

\begin{equation}
N = C \exp\left(-\frac{EW^2}{2\sigma_{\textrm{gauss}}^2}\right)
\label{eq:gauss}
\end{equation}
Here, $N$ is the number of LAEs per EW bin, $C$ is a normalization constant from the fit, $EW$ is the rest-frame Ly$\alpha$ equivalent width ($EW_{\textrm{obs}}$/(1 + $z$)), and $w_0$ and $\sigma_{\textrm{gauss}}$ are the characteristic scale parameters of the exponential and Gaussian distributions, respectively. Applying these fits to the EW distributions of the \textit{SE LAE Catalog} and \textit{Tr LAE Catalog} (Figure \ref{fig:EW Dist}) reveals two markedly different distributions, with the corresponding fit parameters listed in Table \ref{tab:EW Fits}.

\begin{table}[ht]
\centering
\begin{tabular}{llccc}
\hline
Sample (Size) & Parameter & $\chi_\nu^2$ & $f_{\rm cont}$ \\
\hline\hline
\multirow{2}{*}{SE (5354)} & $w_0 = 67.5 \pm 1\,\mathrm{\AA}$ & 2.05 & 0.15 \\    & $\sigma = 91.0 \pm 1\,\mathrm{\AA}$ & 4.22 & 0.253 \\
\hline
\multirow{2}{*}{Tr (12521)} & $w_0 = 41.4 \pm 0.6\,\mathrm{\AA}$ & 9.23 & 0.268 \\    & $\sigma = 48.7 \pm 0.5\,\mathrm{\AA}$ & 14.5 & 0.364 \\
\hline
\multirow{2}{*}{\shortstack[l]{Tr w/o Tr-only\\LAEs (6135)}} & $w_0 = 67.7 \pm 1\,\mathrm{\AA}$ & 3.35 & 0.216 \\    & $\sigma = 90.5 \pm 1\,\mathrm{\AA}$ & 5.92 & 0.322 \\
\hline
\multirow{2}{*}{\shortstack[l]{SE (New Criterion)\\(4853)}} & $w_0 = 72.1 \pm 1\,\mathrm{\AA}$ & 1.63 & 0.137 \\    & $\sigma = 96.7 \pm 1\,\mathrm{\AA}$ & 3.47 & 0.234 \\
\hline
\multirow{2}{*}{\shortstack[l]{Tr (New Criterion)\\(9905)}} & $w_0 = 51.9 \pm 0.8\,\mathrm{\AA}$ & 6.57 & 0.248 \\    & $\sigma = 66.2 \pm 0.8\,\mathrm{\AA}$ & 11.0 & 0.357 \\
\hline
\end{tabular}
\caption{Best-fit parameters for the exponential ($w_0$) and Gaussian ($\sigma_{\mathrm{gauss}}$) fits to the EW distributions of the \textit{SE LAE Catalog}, \textit{Tr LAE Catalog}, and the \textit{Tr LAE Catalog} excluding \textit{Tr-only LAEs}, as well as the \textit{SE LAE Catalog} and \textit{Tr LAE Catalog} after applying our proposed criterion discussed in Section \ref{sec:New LAE Selection Criterion}. The corresponding reduced $\chi^2$ values for each fit are also listed, along with the estimated contamination fraction, assuming the fitted models represents the true intrinsic EW distribution.}
\label{tab:EW Fits}
\end{table}

While neither model provides an exceptional fit, the \textit{Tr LAE Catalog} shows significantly poorer agreement with both the exponential and Gaussian models, driven by a pronounced excess of low-EW sources. Since continuum-only contaminants do not exhibit strong Ly$\alpha$ emission, they are expected to populate the low-EW regime, suggesting that the observed excess arises from contamination by these sources. Assuming the fitted model represents the true intrinsic EW distribution of genuine LAEs, the \textit{Tr LAE Catalog} also exhibits a substantially larger estimated contamination fraction than the \textit{SE LAE Catalog}. Upon further analysis, we find that the \textit{Tr-only LAEs} category is primarily responsible for shifting the EW distribution of the \textit{Tr LAE Catalog} toward lower values. Removing the \textit{Tr-only LAEs} produces an EW distribution that is nearly identical to that of the \textit{SE LAE Catalog}, and correspondingly reduces the estimated contamination fraction. However, despite yielding nearly identical best-fit parameters to the \textit{SE LAE Catalog}, the contamination estimate remains higher, indicating that the \textit{Tr LAE Catalog} without the \textit{Tr-only LAEs} still exhibits a greater positive deviation from the fitted model.

\SectionSpace


\section{Discussion}
\label{sec:Discussion}





\subsection{High-EW Sources} 
\label{sec:Interpretation of EW Distributions}



The differences observed between the SE and Tr LAE samples provide insight into how source detection and photometric methods can influence the properties of selected LAE populations. At the same time, similarities between the two samples can help distinguish features intrinsic to the LAE population from those introduced by the detection method.

We find that both the \textit{SE LAE Catalog} and \textit{Tr LAE Catalog} contain a similar fraction of high-EW sources, with $\sim 11\%$ of each sample exhibiting EW values greater than $240\,\mathrm{\AA}$. LAEs with EW values in this range are either inflated due to noisy measurements caused by spurious narrow-band detections \citep{Firestone2024}, or the result of intrinsically extreme systems, such as young, metal-poor stellar populations \citep{Kashikawa2012}; however, their true nature requires robust spectroscopic measurements and well-understood ISM behavior. 

The agreement of this fraction across both methods, despite their systematic differences in photometry and selection, suggests that this high-EW population is unlikely to be solely a product of the detection technique. This argues against the 11\% being dominated by spurious narrow-band detections which should have different rates in each method. But it remains a question whether EW greater than $240\,\mathrm{\AA}$ is primarily caused by noisy EW measurements or a genuine feature of the underlying LAE population. As a result, this subset of LAEs represents an interesting target for future work aimed at distinguishing between these possibilities.

\subsection{Source Blending and Deblending}
\label{sec: Deblending Sources}

Another possible explanation for the differences between the SE and Tr LAE catalogs is their treatment of blended sources. Because Tr uses forward modeling, it may be better able to distinguish neighboring sources that SE detects as a single blended object. In cases where an LAE is blended with a non-LAE source, the additional continuum emission could reduce the measured EW, potentially causing the LAE to fail our SE selection criteria.

 To investigate whether differences in source blending contribute to the discrepancies between the catalogs, we searched for neighboring sources within 2\,arcseconds using the \textit{Tr Source Catalog}. We identified approximately 1,400 LAEs in the \textit{Tr LAE Catalog} and 500 LAEs in the \textit{SE LAE Catalog} with nearby Tr sources. We then compared the SE EW distributions of LAEs with and without nearby sources to determine whether the presence of neighboring objects systematically affects the measured photometry. We find no clear differences between these distributions, suggesting that nearby sources do not have a significant systematic effect on the measured SE EWs. These results suggest that differences in source blending are unlikely to be a dominant contributor to the discrepancies between the SE and Tr LAE catalogs.

\subsection{New LAE Selection Criterion} 
\label{sec:New LAE Selection Criterion}

One of the primary advantages of narrow-band surveys is their ability to identify high-redshift galaxies whose continua are too faint to be detected reliably in broad-band imaging. At $z \sim 3$, the observed spectrum of a galaxy is strongly affected by neutral hydrogen along the line of sight. Photons with wavelengths shorter than the Lyman limit (rest-frame 91.2\,nm) are absorbed almost completely, while wavelengths between the Lyman limit and Ly$\alpha$ (rest-frame 121.6\,nm) are partially absorbed by intervening hydrogen clouds, producing the Ly$\alpha$ forest. As a result, the continuum blueward of Ly$\alpha$ is significantly suppressed relative to redder wavelengths, creating a sharp spectral break.

For LAEs at $z\sim3.1$, the $u$-band primarily samples the strongly suppressed region near and below the Lyman limit, while the $g$-band primarily samples wavelengths between the Lyman limit and Ly$\alpha$, where the continuum is less strongly suppressed by the Ly$\alpha$ forest. We therefore expect the $u$-band flux to generally be lower than the $g$-band flux. While this provides a physically motivated way to distinguish true high-redshift LAEs from low-redshift interlopers or continuum-dominated sources, care must be taken when incorporating a Lyman-break requirement into the LAE selection process. Requiring a clearly detected Lyman break could exclude genuine LAEs whose continua are too faint to produce a significant $u$-band dropout, thereby reducing one of the primary advantages of narrow-band selection.

\begin{figure}[ht]
    \centering
    \includegraphics[width=\columnwidth]{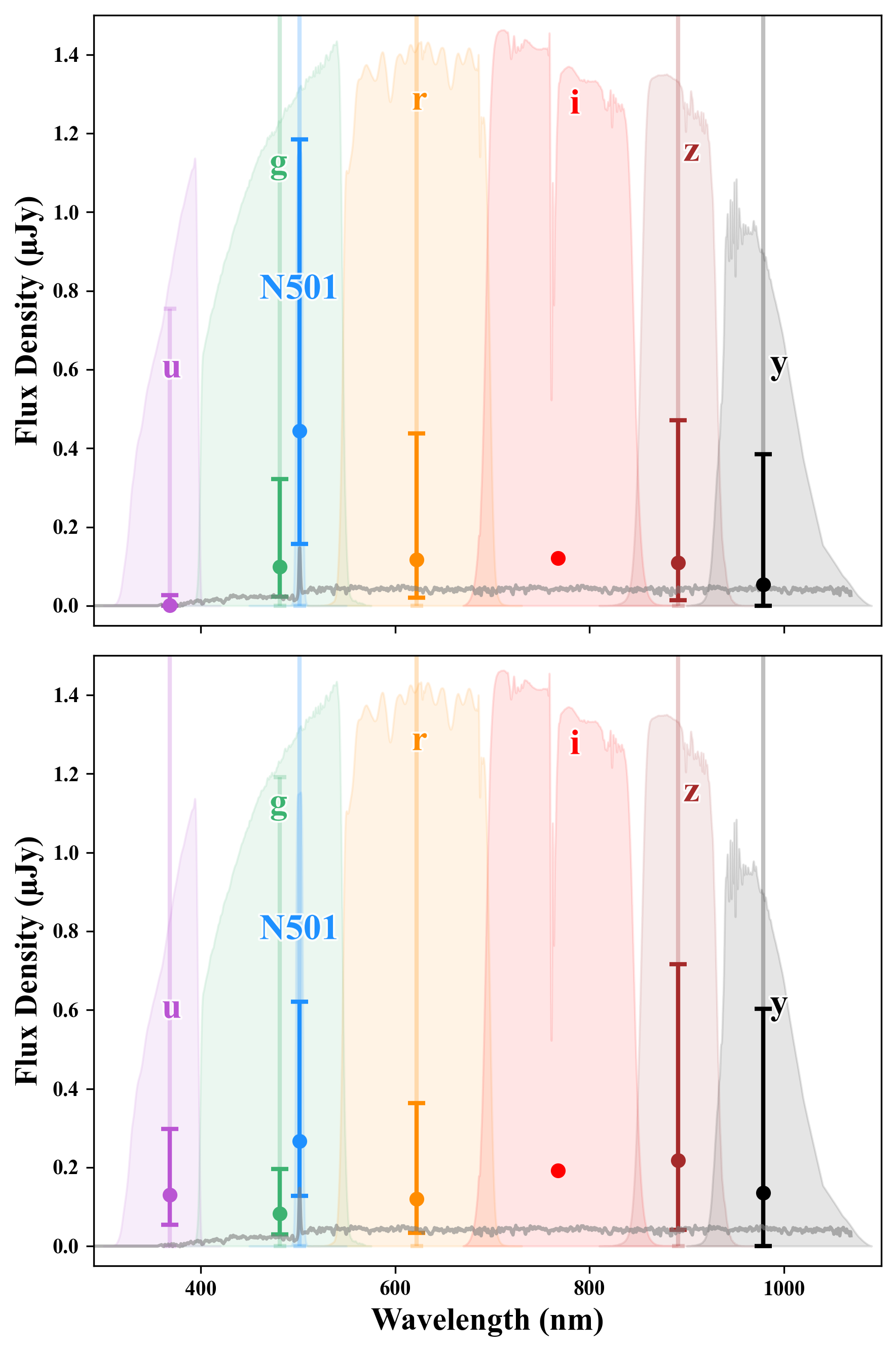}
    \caption{Stacked median spectral energy distributions for \textit{SE+Tr LAEs} with Tr photometry (top) and \textit{Tr-only LAEs} that do not meet the proposed Lyman-break criterion (bottom). Each color represents a different filter. Points show the median flux density value in each filter scaled to the $i$-band. The blue color is for our $N501$ filter while the others represent the broad-band filters ($u$, $g$, $r$, $i$, $z$, and $y$). The more opaque error bar is the 50\% scatter range around the median value and the less opaque is the 95\% scatter range. The gray spectrum \citep{Shapley2003} highlights the expected spectrum of an LAE, including IGM absorption.}
    \label{fig:SEDs}
\end{figure}

Figure \ref{fig:SEDs} presents the stacked median spectral energy distributions of the \textit{SE+Tr LAEs} and a subset of \textit{Tr-only LAEs} that do not have the expected Lyman-break behavior. While the \textit{SE+Tr LAEs} exhibit the expected suppression of the continuum blueward of Ly$\alpha$, these \textit{Tr-only LAEs} display markedly different behavior, including a suspicious depression in the $g$-band relative to the $u$-band. This behavior is inconsistent with the expected broad-band colors of true LAEs and motivates the use of a Lyman-break-based selection criterion.

Rather than requiring that the $g$-band flux simply exceeds the $u$-band flux, we instead require consistency with the expected Lyman break corresponding to $u-g\geq1$, while accounting for the photometric uncertainties. Sources are rejected only if their measured fluxes are inconsistent with this threshold beyond the level expected from photometric noise. We therefore introduce the following selection criterion:

\begin{equation}
0.4\,f_g > f_u - \sqrt{\sigma_g^2 + \sigma_u^2}.
\label{eq:LAE Criterion}
\end{equation}

\begin{figure}[ht]
    \centering
    \includegraphics[width=\columnwidth]{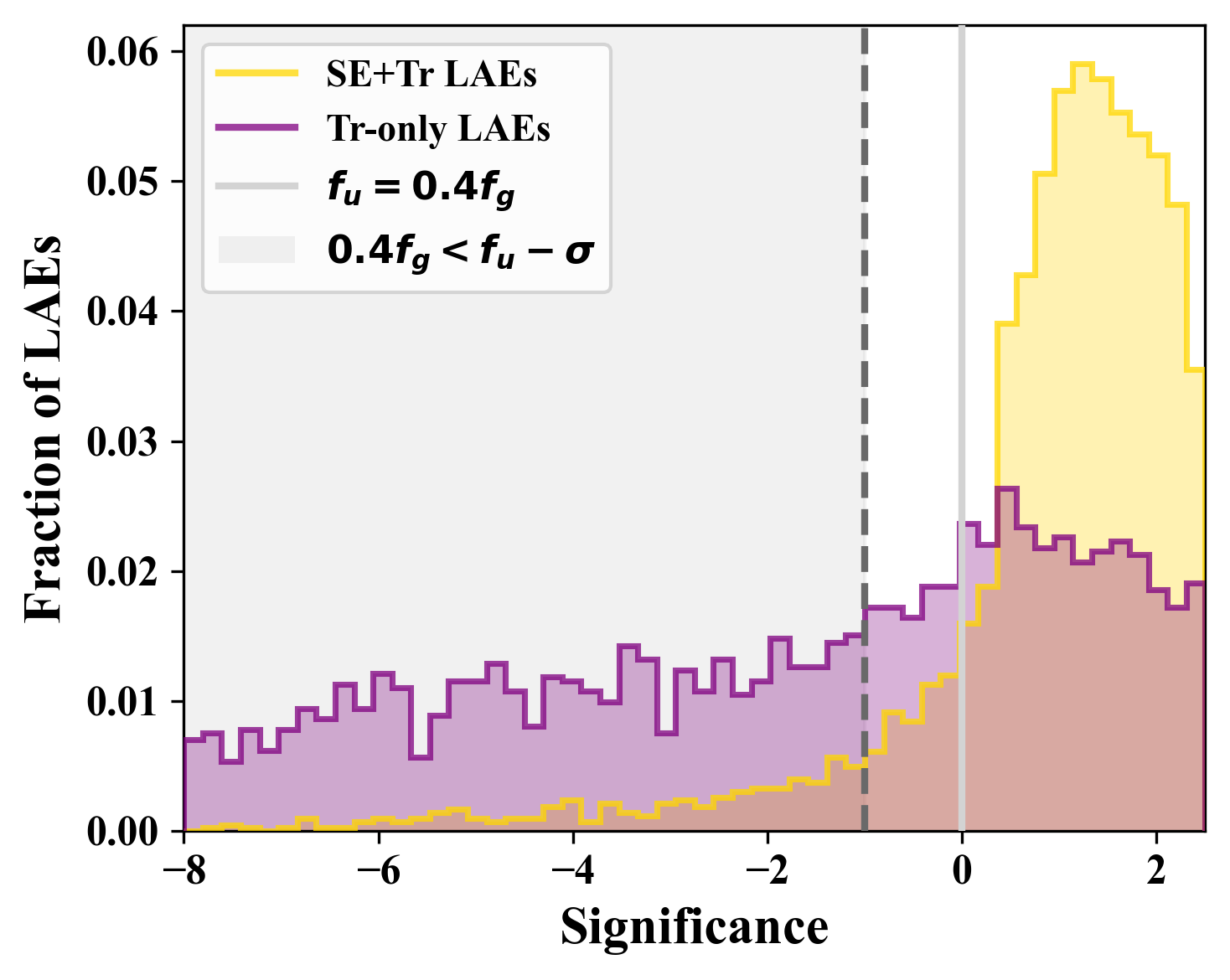}
    \caption{Distribution of the $u$- and $g$-band flux significance, $(0.4f_g-f_u)/\sqrt{\sigma_g^2+\sigma_u^2}$, for the \textit{SE+Tr LAEs} with SE photometry and for \textit{Tr-only LAEs}. The gray shaded region shows sources to the left of the dashed line, which are rejected by the criterion.  Sources between the dashed line and solid line ($0.4f_g=f_u$) are retained because they remain consistent with this threshold within the photometric uncertainties.}
    \label{fig:Color Sig Plot}
\end{figure}

Figure~\ref{fig:Color Sig Plot} illustrates the motivation for this criterion by comparing the distribution of the $u$- and $g$-band significance for the \textit{SE+Tr LAEs} and \textit{Tr-only LAEs}. A strict requirement of $u-g>1$, equivalent to $f_u<0.4f_g$, would reject all sources with negative significance, including many for which the measured fluxes remain consistent with this threshold within the photometric uncertainties. The proposed criterion instead rejects only sources with significance less than $-1$, preserving objects whose apparent deviation from the expected Lyman break is consistent with noise.

Applying this criterion to the full LAE catalogs removes 501 sources from the \textit{SE LAE Catalog} and 2,616 sources from the \textit{Tr LAE Catalog}, 1,818 of which belong to the \textit{Tr-only LAE} category ($\sim\,28$$\%$ of all Tr-only LAEs). This further supports the conclusion that Tr is more prone to including contaminant sources than SE.

The impact of this criterion on the EW distributions is shown in Table~\ref{tab:EW Fits}. For both the \textit{SE LAE Catalog} and \textit{Tr LAE Catalog}, the best-fit exponential ($w_0$) and Gaussian ($\sigma_{\textrm{gauss}}$) parameters increase relative to the original catalogs, indicating the preferential removal of low-EW sources. The quality of both fits also improves, with the reduced $\chi_\nu^2$ values decreasing for both the exponential and Gaussian models. Likewise, the estimated contamination fractions decrease for both catalogs. Although the improvement is more pronounced for the \textit{SE LAE Catalog}, the updated \textit{Tr LAE Catalog} remains more contaminated than the \textit{SE LAE Catalog} and does not achieve the level of agreement obtained by excluding the \textit{Tr-only LAEs}. This suggests that the proposed criterion removes a substantial fraction of contaminant sources while preserving the majority of genuine LAEs, but does not fully account for the excess contamination present in the Tr-selected sample.

\subsection{Application to Other LAE Surveys} 
\label{sec:Application to Other LAE Surveys}

Although this criterion was developed for the ODIN $N501$ survey, the same methodology can be applied to other narrow-band LAE surveys by selecting broad-band filters that lie on either side of the Lyman break at the appropriate redshift. For the ODIN $N673$ survey ($z\sim4.5$), the Lyman break shifts between the $g$- and $r$-bands, making an analogous criterion based on the requirement that $f_r > f_g$ appropriate. As with the $N501$ survey, this comparison should account for the combined photometric uncertainties to avoid rejecting genuine LAEs whose continua are too faint to produce a statistically significant dropout.

For the ODIN $N419$ survey ($z\sim2.4$), the Lyman break lies blueward of the available broad-band coverage and therefore cannot be directly measured. However, the same approach can still be used to identify suspicious sources with unexpectedly high $u$-band fluxes relative to the $g$-band. Although this does not provide a direct Lyman-break test, it offers an additional diagnostic for identifying likely low-redshift interlopers within the LAE sample.

More generally, this method can be applied to any LAE survey with broad-band observations spanning both sides of the expected Lyman break. Surveys with multiple broad-band filters are particularly well suited to this approach because they can use two broad-band filters to estimate the continuum \citep{Firestone2024} while simultaneously performing an independent Lyman-break check. Rather than requiring the detection of a Lyman break, which may exclude continuum-faint LAEs, this approach instead identifies sources that exhibit an obvious lack of the expected Lyman-break signature.

\SectionSpace


\section{Conclusions and Future Work}
\label{sec:Conclusion}

In this paper, we compared SE and Tr source detection methods for the identification and photometric characterization of LAEs in the 9 deg$^2$ ODIN-COSMOS field at $z = 3.1$. Our main conclusions are summarized below.

\begin{enumerate}

    \item Comparison of source catalog photometry shows that SE and Tr generally agree across most flux ranges, though systematic differences emerge at both the high and low-flux extremes. At higher flux densities, Tr tends to measure systematically larger flux densities than SE, while at lower flux levels the detection methods diverge in which sources they detect. These disagreements are most apparent for narrow-band-dominated sources with weak continuum emission, suggesting that differences in source detection sensitivity and uncertainty estimation contribute to the observed asymmetries in LAE selection between the two methods.
    
    \item We find that differences between the SE and Tr LAE catalogs are primarily driven by systematic offsets in narrow-band excess measurements, with Tr generally measuring higher values and selecting a number of LAE candidates that is nearly double the amount of SE. This behavior is driven by differences in continuum estimation, where Tr tends to measure lower flux density in the broad-bands $-$ particularly the $g$ and $r$ filters that are used in estimating narrow-band excess $-$ leading to larger narrow-band excess values.
    
    \item Analysis of equivalent width (EW) distributions further implies that the \textit{Tr LAE Catalog} contains a significant excess of low-EW sources, resulting in poorer agreement with both exponential and Gaussian models. This excess is largely driven by the \textit{Tr-only LAEs} population which makes up half the \textit{Tr LAE Catalog}, indicating that Tr is more prone to including continuum-dominated sources that are not detected by SE. Removing these sources brings the Tr EW distribution into close agreement with that of SE and lowers the overall reduced $\chi^2$ values, indicating a reduction in contamination. However, the estimated continuum fraction still remains larger than that of the \textit{SE LAE Catalog}.
    
    \item We also find that 11$\%$ of both the \textit{SE LAE Catalog} and \textit{Tr LAE Catalog} have EW measurements greater than 240\,\AA, which is the speculated range for LAEs composed of young, metal-poor stellar populations \citep{Kashikawa2012}. This suggests a high EW-tail is not unique to one detection method and may therefore represent a promising population for future study.

    \item We introduce a new selection criterion based on the expected Lyman break behavior of high-redshift galaxies at $z\sim3$, where absorption from intervening hydrogen in the Ly$\alpha$ forest suppresses continuum emission blueward of the Lyman limit, producing a $u-g$ color consistent with a Lyman break. By requiring consistency with this expected color while accounting for the photometric uncertainties, the criterion most strongly impacts the \textit{Tr-only LAEs} category within the \textit{Tr LAE Catalog}, removing $\sim28\%$ of these sources. The resulting improvement in the EW distribution, including lower contamination fractions and better agreement with the fitted models, supports the conclusion that Tr includes a larger fraction of contaminants and demonstrates the usefulness of incorporating expected Lyman break behavior into LAE selection.
    
\end{enumerate}

Taken together, these results indicate that the \textit{SE LAE Catalog} provides a cleaner sample of $z\sim3.1$ LAEs than the \textit{Tr LAE Catalog} for the ODIN survey. Although Tr offers several advantages through its forward-modeling approach, our implementation produces a larger population of continuum-only contaminants that broadens the EW distribution and increases the inferred contamination fraction. While removing the \textit{Tr-only LAEs} and applying our additional Lyman break consistency criterion both improve the Tr-selected sample, the resulting catalog still does not achieve the level of performance obtained with the \textit{SE LAE Catalog}. For this reason, we conclude that SE is the preferred source detection and photometry method for LAE selection in ODIN.

These conclusions are specific to the ODIN imaging and photometric pipeline used in this work and should not be interpreted as demonstrating that SE is generally preferable to Tr for LAE selection. Targeted spectroscopic follow-up of sources will ultimately be necessary to determine their true nature and directly measure the purity and completeness of each LAE sample.

\section{Acknowledgments}
\label{sec:Acknowledgments}

This work is based on observations at Cerro Tololo Inter-American Observatory, NSF's NOIRLab (Prop. ID 2020B-0201; PI: K.-S. Lee), which is managed by the Association of Universities for Research in Astronomy under a cooperative agreement with the National Science Foundation. BT, EG, and NF acknowledge support from NSF grant AST-2206222. NF acknowledges support from the NSF Graduate Research Fellowship Program under Grant No. DGE-2233066.  KSL acknowledges financial support from the National Science Foundation under grant Nos. AST-2206705, AST-2408359, and from the Ross-Lynn Purdue Research Foundations. HSH acknowledges support from the National Research Foundation of Korea
(NRF) funded by the Korea government (MSIT; RS-2026-25482692) and the Global-LAMP Program funded by the Ministry of Education (RS-2023-00301976). LG gratefully acknowledges financial support FONDECYT regular project number 1230591, from ANID - MILENIO - NCN2024\_112, from the ANID BASAL project FB210003.


\bibliographystyle{apj}
\bibliography{library}

\end{document}